\documentclass[aps,floatfix,superscriptaddress,a4paper]{revtex4}
\usepackage{graphicx,amsmath,bbm,mathrsfs,amssymb,psfrag}

\evensidemargin \oddsidemargin
\begin{document}

\author{Laure Gouba\\
The Abdus Salam International Centre for
Theoretical Physics (ICTP),\\
 Strada Costiera 11,
I-34151 Trieste Italy. \\
Email: lgouba@ictp.it}

\title{The Yukawa Model in One Space - One Time Dimensions}

\begin{abstract}
The Yukawa Model is revisited in one space - one time dimensions
in an approach completely different to those available in the literature.
We show that at the classical level it is a constrained system.
We apply the Dirac method of quantization of constrained systems.
Then by means of the bosonization procedure we uniformize the Hamiltonian 
at the quantum level in terms of a pseudo-scalar field and the chiral components of 
a real scalar field. 
\end{abstract}

\maketitle

In 1934 the Japanese physicist, Hideki Yukawa, predicted a new particle which
later became known as the pi meson, or the pion for short \cite{yukawa}.
He considered these pions as the carriers of the force exchanged between two nucleons.
The Yukawa coupling is the coupling  between nucleons and pion that has been generalized 
as any coupling between scalars and fermions. 
In particle physics, Yukawa's interaction or Yukawa coupling,
is an interaction between a scalar field $\phi$ and a Dirac field $\psi$ of the type
$ V = g\bar\psi\phi\psi $  for a scalar field  or $V = g\bar\psi i \gamma_5\phi\psi$ for 
a pseudoscalar field, $g$ is called a Yukawa coupling constant. Recently the scalar Yukawa model 
has been introduced, where the Dirac field is replaced by a complex scalar field 
\cite{abreu, nery, vladimir}.
The Yukawa interaction is also used in the Standard Model to describe the coupling between 
the Higgs field and massless quark and lepton fields (i.e., the fundamental fermion particles).
Several papers about Yukawa models can be found in the literature \cite{feder,hague,li, oscar}.
We are interested in the Yukawa Model in one spatial - one time dimensions that we consider as 
a good testing ground of nonperturbative studies in Yukawa models. We start by considering the model 
at the classical level, then we apply the Dirac method of quantization of constrained systems. 
By means of a bosonization procedure we reformulate the model into a quantum model of scalar fields.

We consider in a Minkowski space-time, the Yukawa model where the coupling is between a 
pseudo-scalar field and a Dirac field. The dynamics of this system is given by the Lagrangian 
density, 
\begin{equation}\label{eq1}
 \mathcal{L} = \frac{1}{2}\partial_\mu\phi\partial^\mu\phi 
 + \frac{i}{2}\bar{\psi}\gamma^\mu\partial_\mu\psi - \frac{i}{2}\partial_\mu\bar\psi\gamma^\mu\psi 
 - g\bar\psi i \gamma_5 \phi \psi - V(\phi)\;,
\end{equation}
where $\phi$ is a pseudo-scalar field, $\psi$ is the Dirac massless field, 
$g$ is the coupling constant, $V$ is the potential 
that for the moment is left arbitrary. We consider the model 
in $1+ 1$ dimensional space-time where the spacetime coordinates indices are $\mu = 0,1$ while spacetime metric is 
$\eta_{\mu\nu} = \textrm{diag}(+,-)$. An implicit choice of units is such that $\hbar = c =1$.  The matrices $\gamma^\mu$ 
define the Clifford Dirac algebra associated to the two-dimensional Minkowski space-time whose representation is 
given by the Pauli matrices 
\begin{equation}
 \sigma^1 = 
 \left(\begin{array}{cc}
0 & +1\\
+1 & 0
\end{array}
\right),\quad
\sigma^2 = 
 \left(\begin{array}{cc}
0 & -i\\
i & 0
\end{array}
\right), \quad 
\sigma^3 = 
 \left(\begin{array}{cc}
+1 & 0\\
0 & -1
\end{array}
\right)\; ,
\end{equation}
and 
\begin{equation}
 \gamma^0 = \sigma^1,\quad 
 \gamma^1 = i\sigma^2, \quad 
 \gamma_5 = \gamma^0\gamma^1 = -\sigma^3 .
\end{equation}
Then in this representation of Dirac, the spinor $\psi$ is split into two components as follows
\begin{equation}
 \psi = \left(\begin{array}{c}\psi_+\\
               \psi_-
              \end{array}
\right), \quad 
\psi^\dagger = (\psi_+^\dagger \; \psi_-^\dagger)\;,
\quad \textrm{with}\quad \gamma_5\psi_\pm  = \mp\psi_\pm, \quad \bar\psi = \psi^\dagger\gamma^0
\end{equation}
where $\psi_+$ is the left chirality spinor and $\psi_-$ the right chirality spinor .

We consider a space-time topology $\tau$ that is cylindrical by compactifying the spatial real line 
into a circle $S^1$ of circumference $L = 2\pi R$, where $R$ is the radius of the circle. We 
formally set 
\begin{equation}
 \tau = \mathbb{R}\times S^1\; ,
\end{equation}
with the above topology, it is necessary to define the periodic boundary conditions. We choose 
the following boundary conditions 
\begin{equation}
 \psi_\pm(t,x+L) = -e^{2i\pi\alpha_\pm}\psi_\pm(t,x), \quad 
 t\in \mathbb{R},\; x\in S^1,
\end{equation}
with $\alpha_\pm $ being real constants.

The dimensions of the fields follows from the corresponding kinetic energy 
terms in the Lagrangian. In two dimensional spacetime ($D = 2$), the dimensionality 
of the fields are set 
\begin{equation}
 [\phi] : \frac{D-2}{2} = 0\,;\quad [\psi] : \frac{D-1}{2} = \frac{1}{2}\,.
\end{equation}
The physical dimension of the coupling constant $g$ is determined 
by using ordinary dimensional analysis as follows 
\begin{equation}
 [g]: D - [\phi]-2 [\psi] = \frac{-D + 4 }{2}= 1\,. 
\end{equation}

Let's first determine the equation of motion for each variable. Given the Lagrangian density 
$\mathcal{L}$  and a degree of freedom $X$, the Euler Lagrange equations are determined by
\begin{equation}
 \partial_\mu\frac{\partial\mathcal{L}}{\partial(\partial_\mu X)} = 
 \frac{\partial\mathcal{L}}{\partial X}\;.
\end{equation}
The explicit expression of the Lagrangian density in equation (\ref{eq1}) is given by
\begin{equation}\label{eqq2}
 \mathcal{L} = \frac{1}{2}(\partial_0\phi)^2 - \frac{1}{2}(\partial_1\phi)^2  + 
 \frac{i}{2}\psi^\dagger\partial_0\psi - \frac{i}{2}\partial_0\psi^\dagger\psi 
 + \frac{i}{2}\psi^\dagger \gamma_5\partial_1\psi -\frac{i}{2}\partial_1\psi^\dagger\gamma_5\psi 
 -i g \phi\psi^\dagger \gamma^1\psi - V(\phi).
\end{equation}
The fundamental degrees of freedom are the following:
\begin{equation}
 \phi,\; \psi,\; \psi^\dagger.
\end{equation}
The equation of motion for the variable $\phi$ is the following 
\begin{equation}
 \partial_0\frac{\partial\mathcal{L}}{\partial(\partial_0\phi)} 
 +\partial_1\frac{\partial\mathcal{L}}{\partial(\partial_1\phi)} 
 = \frac{\partial\mathcal{L}}{\partial\phi}\;,
\end{equation}
that is 
\begin{equation}
 (\partial_0^2 -\partial_1^2)\phi + ig\psi^\dagger\gamma^1\psi + \frac{\partial V}{\partial\phi} = 0\;.
\end{equation}
For the variable $\psi$, the equation of motion is 
\begin{equation}
 \partial_0\frac{\partial\mathcal{L}}{\partial(\partial_0\psi)} 
 +\partial_1\frac{\partial\mathcal{L}}{\partial(\partial_1\psi)} 
 = \frac{\partial\mathcal{L}}{\partial\psi}\;,
\end{equation}
that is 
\begin{equation}
 \partial_0\psi^\dagger + \partial_1\psi^\dagger\gamma_5 + 2 g\phi\psi^\dagger\gamma^1 = 0.
\end{equation}
The equation of motion for the variable $\psi^\dagger$, is determine by 
\begin{equation}
 \partial_0\frac{\partial\mathcal{L}}{\partial(\partial_0\psi^\dagger)} 
 +\partial_1\frac{\partial\mathcal{L}}{\partial(\partial_1\psi^\dagger)} 
 = \frac{\partial\mathcal{L}}{\partial\psi^\dagger}\;,
\end{equation}
that is 
\begin{equation}
\partial_0\psi + \partial_1\gamma_5\psi - 2g\phi\gamma^1\psi = 0\;.
\end{equation}

This model is characterized by the existence of constraints that appear naturally from 
the expressions of the conjugate momenta of the degrees of freedom of the system.
The Literature about constrained systems is wide, for more details, one can read 
for instance in \cite{jan}. The momenta variables associated to $\phi,\; \psi,\; \psi^\dagger$ are respectively 
\begin{equation}
 \pi_\phi = \frac{\partial\mathcal{L}}{\partial(\partial_0\phi)} = \partial_0\phi;\quad 
 \pi_\psi = \frac{\partial\mathcal{L}}{\partial(\partial_0\psi)} = -\frac{i}{2}\psi^\dagger; \quad 
 \pi_{\psi^\dagger} = \frac{\partial\mathcal{L}}{\partial(\partial_0\psi^\dagger)} = 
 -\frac{i}{2}\psi \;,
\end{equation}
where the left derivation convention has been performed for the fermionic variables $\psi$ and $\psi^\dagger$ 
that are Grassmann odd variables. The phase space is then characterized by the pairs 
\begin{equation}\label{ps}
 \{ (\phi(t,x), \pi_\phi(t,x)),\; (\psi(t,x),\pi_{\psi}(t,x)),\;  (\psi^\dagger(t,x),\pi_{\psi^\dagger}(t,x)) \}\;.
\end{equation}
By definition, these pairs are canonically conjugated.  In other words their elementary Poisson brackets 
at equal time are given by 
\begin{eqnarray}\label{pb1}
 \{\phi(t,x), \pi_\phi(t,y)\} &=& \delta(x-y) = - \{\pi_\phi(t,y), \phi(t,x)\}; \\\label{pb2}
 \{\psi(t,x), \pi_{\psi}(t,y)\} &=& -\delta (x-y) = \{\pi_\psi (t,y), \psi(t,x)\}; \\\label{pb3}
 \{\psi^\dagger (t,x); \pi_{\psi^\dagger}(t,y)\} &=& -\delta (x-y) = \{\pi_{\psi^\dagger }(t,y), \psi^\dagger(t,x)\}\; .
\end{eqnarray}
Without any confusion and ambiguity we choose to omit in the rest of the paper the variables $(t,x)$.
Now we apply the canonical formalism for quantizing theories with constraints 
(Dirac formalism). This formalism has been successfully used in \cite{lau} and widely in the literature, 
for instance in \cite{jan,klau,das}.

The conjugate momenta $\pi_\psi$ and $\pi_{\psi^\dagger} $ induce the following constraints
\begin{equation}
 \sigma_1  = \pi_\psi + \frac{i}{2}\psi^\dagger; \quad \sigma_2 = \pi_{\psi^\dagger} + \frac{i}{2}\psi\; .
\end{equation}
These constraints are space-time classical configurations that are in terms of the degres of freedom 
of the system. These constraints are called the primary constraints. 
Since the dynamics of the system depends on the primary constraints, it is compulsory to study 
the dynamical evolution of the system. Given the Lagrangian density in (\ref{eqq2}), the canonical Hamiltonian 
density follows as 
\begin{equation}\label{heq}
 \mathcal{H}_0 = \partial_0\phi\pi_\phi + \partial_0\psi\pi_\psi + \partial_0\psi^\dagger\pi_{\psi^\dagger} - 
 \mathcal{L}\;, 
\end{equation}
that is after substitution of $\mathcal{L}$ by its expression in (\ref{eqq2}) 
\begin{equation}\label{eq2}
 \mathcal{H}_0  = \frac{1}{2}\pi_\phi^2 + \frac{1}{2}(\partial_1\phi)^2 -\frac{i}{2}\psi^\dagger\gamma_5\partial_1\psi 
 + \frac{i}{2}\partial_1\psi^\dagger\gamma_5\psi + ig\phi\psi^\dagger \gamma^1\psi + V(\phi)
\end{equation}
and the canonical Hamitonian is 
\begin{equation}
 H = \int dx \mathcal{H}_0.
\end{equation}

Once we have specified the phase space in equation (\ref{ps}), the fundamental Poisson brackets 
in equations (\ref{pb1}), (\ref{pb2}), (\ref{pb3}) and the canonical Hamiltonian density in 
equation (\ref{eq2}), we can study now the evolution of the constraints in order to check if they 
generate other constraints and proceed to their classification according to the Dirac formalism. 

The primary Hamiltonian is given by the summation of the canonical Hamiltonian and 
a linear combination of the primary constraints as follows 
\begin{equation}
 H_1  = H_0 + \int dx (u_{(1)}\sigma_1 + u_{(2)}\sigma_2).
\end{equation}
In order to check whether the constraints $\sigma_i,\; i = 1, 2 $ generate 
other constraints, we solve the equations $\{\sigma_i,\; H_1\} = 0,\; i = 1, 2$\;.
\begin{eqnarray}
 \{\sigma_1, H_1\} = u_{(2)}\int dx (-\frac{i}{2}\delta (x-y)); \quad 
 \{\sigma_2, H_1\} = u_{(1)}\int dx (-\frac{i}{2})\delta (x-y).
\end{eqnarray}
Solving $\{\sigma_i, H_1\} = 0 $ implies some choices for $u_{(i)},\; i = 1, 2$ and that means 
that the constraints $\sigma_i,\: i =1, 2$ do not generate other constraints. The total number
of constraints for this model is then equal to 2. An algebra of the constraints is as follows
\begin{eqnarray}\label{alg1}
 \{\sigma_1,\;\sigma_1\} = 0;\quad 
 \{\sigma_1,\;\sigma_2\} = -\frac{i}{2}\delta (x-y);\quad 
 \{\sigma_2,\sigma_1\} = -\frac{i}{2}\delta (x-y);\quad 
 \{\sigma_2,\;\sigma_2\} = 0.
\end{eqnarray}
With the algebra in (\ref{alg1}), we conclude that all the constraints are of second class.
Let's call $\Delta$ the matrix of the Poisson brackets of the second class constraints.
According to the Dirac formalism of quantization, we should define the algebra of the 
Dirac brackets by using the general formula 
\begin{equation}
 \{ f, \; g \}_D=  \{ f,\; g \}  - \sum_{s,s'}\{ f,\;\sigma_s \}C^{ss'}\{ \sigma_{s'},\;g \}, 
\end{equation}
where $f$ and $g$ are two degrees of freedom, $\sigma_s,\; \sigma_{s'}$ the constraints and  $C$ the inverse matrix 
of the matrix $\Delta$. 
Then it follows for our system the Dirac brackets 
\begin{eqnarray}
 \{\phi, \pi_\phi\}_D &=& \delta (x-y) = -\{\pi_\phi,\phi\}_D;\\
 \{\psi_{\pm},\psi_{\pm}^\dagger\}_D &=& -i\delta (x-y) = \{\psi_{\pm}^\dagger, \psi_\pm\}_D .
\end{eqnarray}

The fundamental Hamiltonian formulation of the system is then given after the complete 
analysis of the system by the degrees of freedom $\phi(t,x),\; \pi_\phi(t,x),\; \psi(t,x), \;\pi_{\psi}(t,x), \;
\psi^\dagger(t,x), \;\pi_{\psi^\dagger}(t,x)$, the fundamental symplectic structure is given by 
the Dirac brackets, that appear now implicitly since we omit the index $D$ as 
\begin{eqnarray}
 \{\phi(t,x), \pi_\phi(t,y)\} &=& \delta (x-y) = -\{\pi_\phi(t,x),\phi(t,y)\};\\
 \{\psi_{\pm}(t,x),\psi_{\pm}^\dagger(t,y)\} &=& -i\delta (x-y) = \{\psi_{\pm}^\dagger(t,x), \psi_\pm(t,y)\},
\end{eqnarray}
and the fundamental Hamiltonian density
\begin{equation}\label{eq3}
 \mathcal{H}  = \frac{1}{2}\pi_\phi^2 + \frac{1}{2}(\partial_1\phi)^2 -\frac{i}{2}\psi^\dagger\gamma_5\partial_1\psi 
 + \frac{i}{2}\partial_1\psi^\dagger\gamma_5\psi + ig\phi\psi^\dagger \gamma^1\psi + V(\phi).
\end{equation}

Now we proceed by canonical quantization, that is the correspondence principle that states that 
to each of the classical structures should correspond a similar structure for the quantum system. Then 
at the quantum level, the phase space is the abstract Hilbert space which elements are called 
quantum states. To the classical variables of the phase space correspond linear operators acting 
on the Hilbert space. To the Poisson brackets correspond now an algebraic structure of commutation relations 
for the quantum system. We still consider that $\hbar = 1 = c$. 
We have then the fundamental bosonic and fermionic operators: 
\begin{equation}
 \hat\phi(t,x),\; \hat\pi_\phi(t,x),\; \hat\psi_\pm(t,x),\; \hat \psi_\pm^\dagger(t,x),
\end{equation}
that satisfy the fundamental commutation and anticommutation relations respectively for the bosonic 
operators and the fermionic operators: 
\begin{eqnarray}
 [\hat\phi(t,x),\; \hat\pi_\phi(t,y)] &=& i\delta (x-y) = - [\hat\pi_\phi(t,x),\; \hat\phi(t,y)]\; ;\\
 \{\hat\psi_\pm(t,x),\; \hat\psi_\pm^\dagger(t,y)\} &=& \delta (x-y) = \{\hat\psi_\pm^\dagger(t,x),\;\hat\psi_\pm (t,y)\}\;.
\end{eqnarray}
The quantum Hamiltonian density is given by
\begin{eqnarray}\label{ham1}
 \hat{\mathcal{H}} = \frac{1}{2}\hat\pi^2_\phi + \frac{1}{2}(\partial_1\hat\phi)^2 -\frac{i}{2}\hat\psi^\dagger\gamma_5\partial_1\hat\psi
 + \frac{i}{2}\partial_1\hat\psi^\dagger\gamma_5\hat\psi
 + ig\hat\phi\psi^\dagger\gamma^1\hat \psi + V(\hat\phi).
\end{eqnarray}
The Hamiltonian (\ref{ham1}) can be also written in terms of their chiral components of the fermionic operators as 
\begin{eqnarray}\label{ham2}\nonumber
 \hat{\mathcal{H}} &=&  \frac{1}{2}\hat\pi_\phi^2 + \frac{1}{2}(\partial_1\hat\phi)^2 + 
 \frac{i}{2}\left(\hat\psi_+^\dagger\partial_1\hat\psi_+ - \partial_1\hat\psi_+^\dagger\hat\psi_+  \right)
 -\frac{i}{2}\left(\hat\psi_-^\dagger\partial_1\hat\psi_- - \partial_1\hat\psi_-^\dagger\hat\psi_-  \right)\\
 &+& i g\hat\phi\left(\hat\psi_+^\dagger\hat\psi_- - \hat\psi_-^\dagger\hat\psi_+  \right) + V(\hat\phi).
\end{eqnarray}

It is well known that in $1+ 1$ spacetime dimensions, the fermionic operators can be 
expressed in terms of bosonic operators by means of vertex operators and the Klein factors.
This procedure is called bosonization. The inverse procedure called fermionization also exists 
but is less used in the literature \cite{lgo}. The aim of this paper is to uniformize the quantum 
representation of the system, we choose to bosonize the fermionic operators in order to uniformize 
the Hamiltonian in terms of only bosonic operators. 
We consider the Schr\"odinger picture where the time variable is fixed and that we choose equal 
to zero. The notation $:\;\;:$ expresses the normal ordering in which the creation operators 
should be placed at the left of the annihilation operators. We bosonize then the chiral fermionic operators in terms 
of chiral bosonic operators. This procedure has been already well done in \cite{ga}, \cite{avos}.
Referring then to the results in \cite{ga} and \cite{avos}, the chiral components of the fermionic operators 
in (\ref{ham2}) are bosonized as follows
\begin{eqnarray}\label{fb1}
 \hat\psi_\pm (x) &=& \frac{1}{L} e^{\pm\frac{i\pi}{L}x}e^{i\rho_\pm\frac{\pi}{2}\hat{p}_\mp} 
 : e^{\pm i\lambda\hat\varphi_\pm(x)}:\; ;\\\label{fb2}
 \hat\psi_\pm^\dagger (x) &=& \frac{1}{L} e^{\pm\frac{i\pi}{L}x}e^{-i\rho_\pm\frac{\pi}{2}\hat{p}_\mp}
 : e^{\mp i\lambda\hat\varphi_\pm(x)}:\;,
\end{eqnarray}
where 
\begin{eqnarray}\label{fb3}
 \hat\varphi_\pm(x) = \hat q_\pm \pm \frac{2\pi}{L}\hat{p}_\pm\; x  + 
 \sum_{n = 1}^{+\infty}\frac{1}{\sqrt{n}}\left(
 a_{\pm,n}^\dagger e^{\pm\frac{2 i \pi}{L}nx} + 
 a_{\pm,n} e^{\mp\frac{ 2 i \pi}{L}nx} \right),
\end{eqnarray}
are the chiral components of a real scalar bosonic field $\hat\varphi = \hat\varphi_+ + \hat\varphi_-$. The parameters 
$\lambda$ and $\rho_\pm$ are such that $\lambda = \pm 1$ and $\rho_\pm^2 = 1 = \lambda^2$. The Klein factor is given by
$e^{i\rho_\pm\frac{\pi}{2}\hat{p}_\mp}$.
We have 
\begin{equation}
 \partial_1\hat\varphi_\pm (x) = \pm\frac{2i\pi}{L}\left(p_\pm + 
 i\sum_{n=1}^{+\infty}\sqrt{n}\left(a_{\pm,n}^\dagger e^{\pm\frac{2 i \pi}{L}nx} - 
 a_{\pm,n}e^{\mp\frac{2 i \pi}{L}nx}\right)\right),
\end{equation}
and the algebra of the bosonic representation is 
\begin{equation}
\left[\hat q_\pm,\; \hat p_\pm\right] = i ;\quad
\left[a_{\pm,n},\; a_{\pm,m}^\dagger\right] = \delta_{nm}, \quad n,\;m \ge 1\; ,
\end{equation}
\begin{equation}
 \left[ \hat\varphi_\pm(x),\;\partial_1\hat\varphi_\pm(y)\right] = \pm 2i\pi\delta (x-y).
\end{equation}
For the normal ordering, the convention is that the operators 
$\left(\begin{array}{c}\hat q_\pm \\
a^\dagger_{\pm,n}
\end{array}\right)$ 
should be placed at the left of the operators
$\left(\begin{array}{c}\hat p_\pm \\
a_{\pm,n}
\end{array}\right)$. 
Using the procedure of point splitting, that is necessary for the well definition of 
the composites fermionic operators, and the Baker-Campbell-Hausdorf formula, we show that 
\begin{eqnarray}
 \hat\psi^\dagger_+\partial_1\hat\psi_+  - \partial_1\hat\psi^\dagger_+\hat\psi_+ &=& 
 (-i)\left(\frac{1}{\pi}(\partial_1\hat\varphi_+)^2 - \frac{\pi}{L^2}\right);\\
 \hat\psi^\dagger_-\partial_1\hat\psi_-  - \partial_1\hat\psi^\dagger_-\psi_- &=& 
 (i)\left(\frac{1}{\pi}(\partial_1\hat\varphi_-)^2 - \frac{\pi}{L^2}\right);\\
 \hat\psi^\dagger_+\hat\psi_-  - \hat\psi^\dagger_-\hat\psi_+ &=& 
 -\frac{2i}{L} :\sin [\frac{\pi}{2}(\rho_+\hat p_- - \rho_-\hat p_+) +\lambda(\hat\varphi_+ +\hat\varphi_-]:\;.
\end{eqnarray}
We set now $\lambda = 1$ and the uniformized quantum Hamiltonian density of the 1+1 Yukawa model is given by
\begin{eqnarray}\nonumber
 \hat{\mathcal{H}} &=&  : \frac{1}{2}\hat\pi_\phi^2 + \frac{1}{2}(\partial_1\hat\phi)^2 + \frac{1}{2\pi}(\partial_1\hat\varphi_+)^2 
 + \frac{1}{2\pi}(\partial_1\hat\varphi_-)^2 
 + \frac{2}{L}g\,\hat\phi\, \sin \left(\frac{\pi}{2}(\rho_+\hat{p}_- -\rho_-\hat{p}_+) + \hat\varphi_+ +\hat\varphi_-\right) + V(\hat\phi): \\\label{eqf}
 &-& \frac{\pi}{L^2},
\end{eqnarray}
where $\hat\phi$ is the quantum pseudo-scalar field in (\ref{ham1}) and $\hat\varphi_\pm$ the chiral components of 
the real scalar field $\hat\varphi$ in equations (\ref{fb1}), (\ref{fb2}), (\ref{fb3}). 
The quantity $\frac{\pi}{L^2}$ can be interpreted as the Casimir Energy.

As concluding remarks, we can first notice the absence of first class constraints at 
the classical level, means there are no gauge symmetry generators, that makes the model more simple. 
We did not discuss about the symmetries and conserved charges. The potential is left arbitrary, a nice 
choice would be the Higgs potential. Some extensions of this work can be performed starting with the equation 
(\ref{eqf}). For instance, the coupling constant $g$ has the dimension of mass, thus setting a mass scale. 
It would be interesting to understand how this mass scale, $g$, determines finally the mass spectrum of the 
different (pseudo) scalar fields in equation (\ref{eqf}).

{\bf Acknowledgments}

L. Gouba is supported by the Abdus Salam International Centre for Theoretical Physics (ICTP), Trieste, Italy.

\end{document}